\documentclass[useAMS,usenatbib]{mn2e}

\usepackage{amssymb,amsmath}
\usepackage{graphicx}
\usepackage{float}
\usepackage{multirow}
\usepackage{aas_macros}
\usepackage{fixltx2e}
\usepackage[usenames]{color}

\newcommand\msun {M$_{\odot}$}
\def\flx{erg cm$^{-2}$ s$^{-1}$}
\def\lum{erg s$^{-1}$}
\def\ctr{cts s$^{-1}$}

\def\chan{{\it Chandra}}

\def\hst{{\it HST}}
\def\cxo{CXO J1225}

\def\g{{\it g'}}
\def\z{{\it z'}}
\def\lx{$L_{\mathrm{X}}$}

\title[A second outbursting HLX]{Discovery of a second outbursting hyperluminous X-ray source}
\author[M. Heida et al.]
{M. Heida$^{1,2}$, P. G. Jonker$^{1,2}$, M. A. P. Torres$^{1,2,3}$\\
$^1$SRON Netherlands Institute for Space Research, Sorbonnelaan 2, 3584 CA Utrecht, the Netherlands\\
$^2$Department of Astrophysics/IMAPP, Radboud University Nijmegen, P.O. Box 9010, 6500 GL Nijmegen, The Netherlands\\
$^3$European Southern Observatory, Alonso de Cordova 3107, Casilla 19001, Vitacura, Santiago 19, Chile\\
}

\begin{document}

\maketitle

\begin{abstract}
We report on six \chan{} and one \hst/WFC3 observation of CXO~J122518.6+144545, discovered by \citet{jonker10} as a candidate hyperluminous X-ray source (HLX), X-ray bright supernova or recoiling supermassive black hole at \lx{} $= 2.2 \times 10^{41}$ \lum{} (if associated with the galaxy at 182 Mpc). We detect a new outburst of the source in a \chan{} image obtained on Nov 20, 2014 and show that the X-ray count rate varies by a factor $> 60$. New \hst/WFC3 observations obtained in 2014 show that the optical counterpart is still visible at \g{} = $27.1 \pm 0.1$, $1 \pm 0.1$ magnitude fainter than in the discovery \hst/ACS observation from 2003. This optical variability strongly suggests that the optical and X-ray source are related. Furthermore, these properties strongly favour an HLX nature of the source over the alternative scenarios. We therefore conclude that CXO~J122518.6+144545 is most likely an outbursting HLX. It is only the second such object to be discovered, after HLX-1 in ESO 243-49. Its high X-ray luminosity makes it a strong candidate to host an intermediate mass black hole.   
\end{abstract}


\section{Introduction}
Intermediate mass black holes (IMBHs) are defined as objects with masses in between those of stellar mass and supermassive black holes (BHs; we use the mass range $10^2 - 10^5$ \msun for IMBHs; for a review see \citealt{vandermarel04}). In the $\lambda$ cold dark matter cosmology with hierarchical structure formation IMBHs are important as possible seeds for supermassive BHs (SMBHs; \citealt{ebisuzaki01}). Scenarios to form IMBHs include direct collapse of gas in atomic cooling haloes in the early Universe (e.g.~\citealt{begelman06,ferrara14}), the collapse of extremely massive Pop III stars (e.g.~\citealt{madau01}) and merging of stars in the nuclei of dense star clusters (e.g.~\citealt{portegieszwart02}). 

However, observational evidence for the existence of IMBHs is scarce. Dynamical studies searching for IMBHs in the centers of globular clusters have yielded inconsistent results (\citealt{lutzgendorf11,lanzoni13}). Radio and X-ray observations have yielded no evidence for accreting IMBHs in globular clusters (\citealt{strader12,haggard13}). There is evidence for the presence of IMBHs with masses around $10^5$ \msun{} in the nuclei of dwarf galaxies (e.g.~\citealt{greene07,reines13}), although these mass determinations depend on extrapolating scaling relations established for SMBHs to lower BH masses.

Ultraluminous X-ray sources (ULXs) are off-nuclear X-ray sources with an X-ray luminosity $> 10^{39}$ \lum{} (see \citealt{feng11} for a review). Assuming isospherical emission and if these sources are Eddington-limited they should contain BHs that are more massive than ordinary stellar mass BHs. Detailed X-ray spectra have revealed that many ULXs are in a so-called ultraluminous state (\citealt{gladstone09,sutton13}) that may be a sign of super-Eddington accretion. These ULXs probably contain stellar mass BHs. For a handful of ULXs there is dynamical evidence for a stellar mass BH (\citealt{liu13,motch14}) and the detection of X-ray pulsations from M82 X-2 proves that that system contains a neutron star (\citealt{bachetti14}). 

However, the most luminous ULXs, often referred to as hyperluminous X-ray sources (HLXs), are still strong candidates to host IMBHs. The best example is HLX-1 in ESO 243-49 (\citealt{farrell09,webb10}). This is the brightest HLX known to date; it shows outbursts reminiscent of those in Galactic X-ray binaries, but with a peak luminosity of $10^{42}$ \lum{} (\citealt{godet09,servillat11}). This luminosity is hard to explain if the accretor is a stellar mass BH (but see \citealt{king14,lasota15}). The outbursts are possibly related to a donor star in a very eccentric orbit that transfers mass when it passes pericentre (\citealt{lasota11,godet14}). 
Other HLXs containing candidate IMBHs include M82-X1 (peak X-ray luminosity $\sim 10^{41}$ \lum, \citealt{strohmayer03, kaaret09b, pasham14}) and NGC2276-3c (peak X-ray luminosity $\sim 6\times 10^{40}$ \lum, \citealt{mezcua15}). 

\citet{jonker10} reported the discovery of a luminous off-nuclear X-ray source (CXO~J122518.6+144545; hereafter \cxo) in the galaxy SDSS~J122518.86+144547.7 (hereafter SDSS J1225). The X-ray source was discovered in an archival 5 ks \chan{} observation taken in 2008. A blue optical counterpart was detected in an \hst/ACS image of the field, with \g{} $= 26.4 \pm 0.1$ and \z{} $ > 25.7$. 
Assuming \cxo{} is at the same distance as the galaxy ($z = 0.0445 \pm 0.0001$, 182 Mpc, \citealt{sdssdr5}), its 0.3 -- 8 keV luminosity in this observation was $2.2 \times 10^{41}$ \lum, making it one of the most luminous ULXs known to date. Alternative explanations proposed by \citet{jonker10} were a recoiling SMBH as a result of the merger of two SMBHs or a Type IIn supernova. 

To determine which of these scenarios is correct, we obtained new \chan{} and \hst{} observations of \cxo{} in 2012 and 2014. We describe these observations, the analysis and results in Section 2. In Sections 3 and 4, we discuss the different scenarios and present our conclusions.
All errors quoted in this Letter are 1--$\sigma$ errors unless otherwise specified. We assume a standard cosmology with $H_0 = 73$ km s$^{-1}$ Mpc$^{-1}$.

\section{Observations, analysis and results}
\subsection{\chan}

\begin{table*}
\begin{small}
  \centering
 \caption{Journal of the \chan{}/ACIS-S observations of \cxo{} used in this Letter. The last column represents the unabsorbed flux. The count rate and flux columns list the 95\% confidence interval calculated following \citet{gehrels86}.}\label{tab:cxo-obs}
\begin{tabular}{rlcccll}
 \hline
 Obs.~ID & Date & MJD & Exp. time & Counts & Count rate & Flux (0.5 -- 10 keV)\\
 & (UT) & & (ks) & & ($10^{-4}$ \ctr) & ($10^{-15}$ \flx) \\
  \hline
  8055 & 2008 Feb 19 & 54515.686 & 5.09 & 22 & 30 -- 60 & 33 -- 70\\
  13295 & 2012 Nov 26 & 56257.804 & 9.78 & 0 & 0 -- 3 & 0 -- 4 \\
  16476 & 2014 Apr 28 & 56775.840 & 18.87 & 3 & 0.4 -- 4 &  0.5 -- 5\\
  16477 & 2014 Jul 22 & 56860.855 & 19.85 & 0 & 0 -- 2 & 0 -- 2\\
  15783 & 2014 Nov 20 & 56981.926 & 19.85 & 20 & 7 -- 15 & 8 -- 20\\
  17558 & 2014 Dec 08 & 56999.847 & 19.78 & 4 & 0.7 -- 5 &  0.8 -- 5\\
  17559 & 2014 Dec 15 & 57006.499 & 39.59 & 0 & 0 -- 0.8 & 0 -- 0.9\\
  \hline  
 \end{tabular}
 \end{small}
\end{table*}

We obtained one 10 ks \chan/Advanced CCD Imaging Spectrometer (ACIS) (\citealt{garmire03}) observation of \cxo{} in 2012 and five (4 $\times$ 20 ks and one 40 ks observation) in 2014 (see Table \ref{tab:cxo-obs}). These observations were done using ACIS-S in very faint mode. We reprocess the events with calibrations available in CALDB version 4.6.7, using version 4.7 of the \chan{} X-ray centre {\sc ciao} tools (\citealt{fruscione06}). In all observations, we extract the source counts in a circular region with radius $2''$. For the background subtraction region we use an annulus centred on the source position with an inner radius of $5''$ and an outer radius of $20''$. 

We detect the source in the observations with Obs.~ID 16476, 15783 and 17558. In the observations with Obs.~ID 13295, 16477 and 17559 the source is not detected (zero counts). We calculate 95\% confidence limits for the count rates in these observations following \citet{gehrels86} --- for comparison we also calculated limits following \citet{kraft91}, but we find that the differences are negligible. The count rates of the detections and upper limits of the non-detections, including the original detection in archival data from 2008, are listed in Table \ref{tab:cxo-obs}. 

Of the 20 counts in observation 15783, several are detected above 2 keV. We use {\sc Xspec} version 12.8.2 and C-statistics (cstat; \citealt{cash79}) modified to account for the subtraction of background counts, the so called W-statistics\footnote{see http://heasarc.gsfc.nasa.gov/docs/xanadu/xspec/manual/}, to fit models to the data. We find that both a power law and a disc blackbody can describe the data. For the power law, we find a photon index $\Gamma = 1.4 \pm 0.5$, with a cstat value of 19.8 using 19 bins and 17 degrees of freedom. For a disc blackbody we find a temperature of $0.6 \pm 0.1$ keV, with a cstat value of 21.8 using 19 bins and 17 degrees of freedom.
To ease comparison with \citet{jonker10}, we assume a standard power law with photon index $\Gamma = 1.7$, a Galactic hydrogen column density $N_\textrm{H} = 2.8 \times 10^{20}$ cm$^{-2}$ (\citealt{dickey90}) and no local absorption to describe the spectrum.

\begin{figure}
\includegraphics[width=0.5\textwidth]{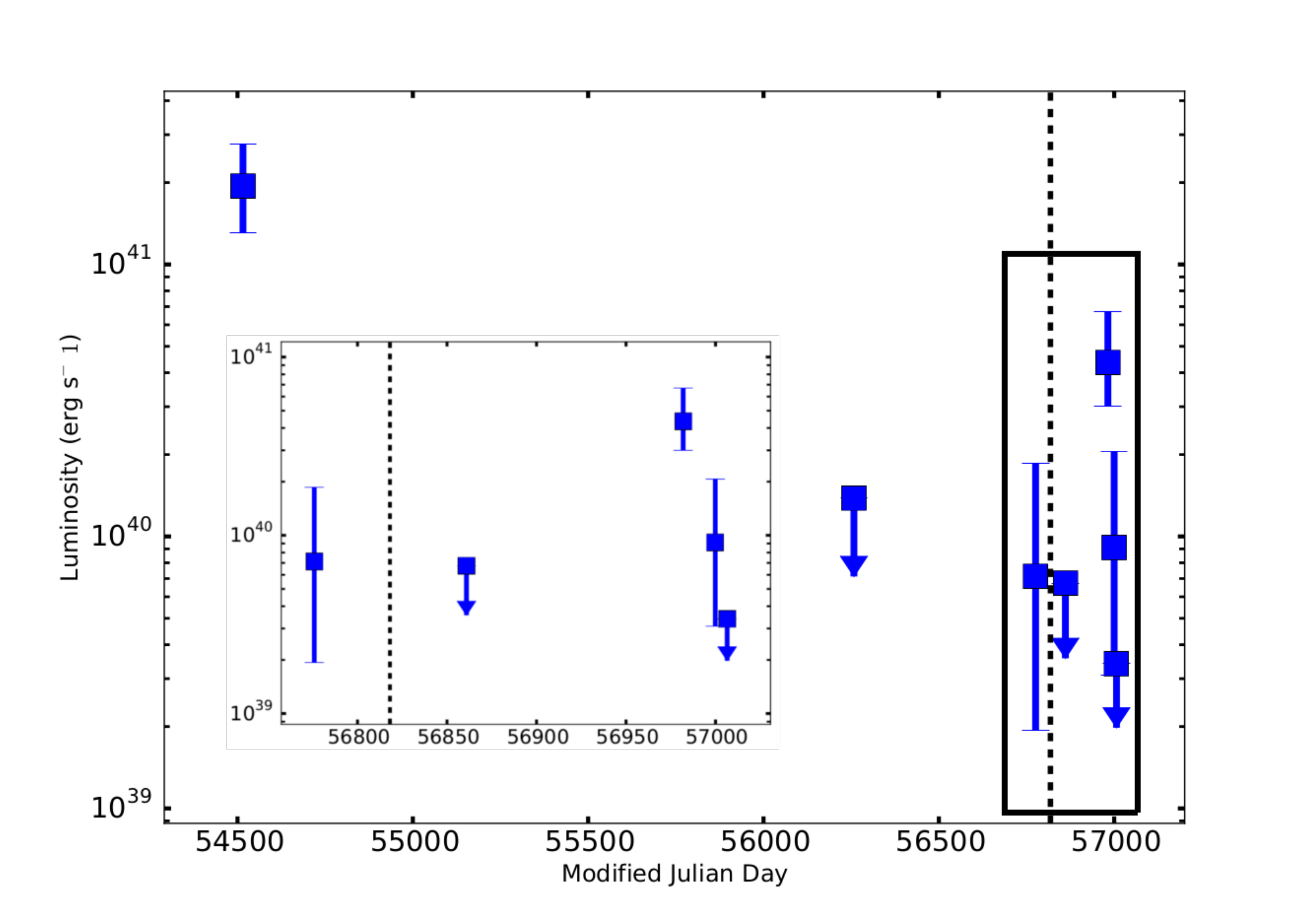}
\caption{The 2008-2014 X-ray light-curve of \cxo. Blue squares are \chan{} observations (detections with 95\% confidence intervals and 95\% confidence upper limits), the black dashed line indicates the date of our \hst/WFC3 observation. The inset is zoomed-in on the 2014 data points, showing the decay of the outburst detected in November.}\label{fig:lc}
\end{figure}

We convert the detected count rates to (unabsorbed) fluxes in the 0.5-10 keV range using {\sc webpimms}, the web version of {\sc Pimms} (\citealt{mukai93}).  The confidence intervals listed in Table \ref{tab:cxo-obs} only take into account the uncertainties in the count rates, not the (unknown) additional uncertainty introduced by the model.
To convert fluxes to luminosities we assume that the flux is isotropic and that \cxo{} is at the distance of SDSS J1225 (182 Mpc). The highest luminosity that \cxo{} reached in these new observations is $5 \times 10^{40}$ \lum, on 2014 November 20 (see Figure \ref{fig:lc}).

We align and stack all observations, using the point source at Right Ascension (R.A.) = 12:25:08.93, Declination (Dec) =  14:46:01.04 in observation 17559 for alignment, to search for X-ray emission from the nucleus of SDSS J1225. The total exposure time of the stacked image is 127.72 ks. In this deep image we detect eight counts in a circle with a radius of $1''$ centred on the position of the nucleus of SDSS J1225 (R.A. = 12:25:18.86, Dec = 14:45:47.7 [J2000], \citealt{sdssdr7}). Assuming a power law with index 1.7 and N$_\textrm{H} = 10^{21}$ cm$^{-2}$ we find an unabsorbed $0.5 - 10$ keV flux of $8 \times 10^{-16}$ \flx{} (absorbed flux = $7 \times 10^{-16}$ \flx). At a distance of 182 Mpc this translates to a luminosity of $3 \times 10^{39}$ \lum.

\subsubsection*{Count rate in quiescence} 
The strongest limit on the count rate of \cxo{} when it is in the low state is set by the 40 ks observation (17559), at $< 8 \times 10^{-5}$ \ctr{} (\lx{} $< 3 \times 10^{39}$ \lum). This is still consistent at the 95\% confidence level with the count rates observed in observations 16476 and 17558, although alternately either one or both of these observations may have caught \cxo{} fading after an outburst, in which case the quiescent level could be (much) lower.

Assuming the count rate is constant over observations 13295, 16476, 16477, 17558 and 17559 at a value so that only occasionally, because of Poisson fluctuations, three or four source counts are detected, we calculate the most likely value of the count rate in the following way. For a range of count rates ($10^{-5} - 5 \times 10^{-4}$ \ctr) we calculate the expected number of source counts in each observation. Then, taking into account the expected number of background counts and assuming both source and background follow a Poisson distribution, we calculate the probability to retrieve the observed number of counts in each observation. For each assumed count rate we then multiply the probabilities found for the five observations. We find that this value is maximal for a count rate of $5 \times 10^{-5}$ \ctr{} (\lx{} $\approx 2.5 \times 10^{39}$ \lum).

If we repeat this calculation but without observation 17558 (as this detection could plausibly be due to a fading outburst) we find a most probable count rate of $2 \times 10^{-5}$ \ctr{} (\lx{} $\approx 10^{39}$ \lum).

\subsection{\hst}

On 2014 June 10 we obtained four dithered \hst/Wide Field Camera 3 (WFC3) observations of \cxo{} using the UVIS detector with the F475W (\g-band) filter (data set ICEF01010). The total exposure time is 2512 s. The optical counterpart that was reported by \citet{jonker10} is clearly visible in the multidrizzled image. 

We use the WFC3 package (version 2.0) of {\sc dolphot} (version 2.0, \citealt{dolphin00}) for the photometric analysis. Following the {\sc dolphot}/WFC3 user guide we run {\sc dolphot} on the flat-fielded, bias corrected \emph{flt} images, excluding the ICEF01ohq frame because there is a cosmic ray hit close to the position of \cxo. We first process the \emph{flt} files with {\it wfc3mask} to mask bad pixels. Then we split the multi-extension fits files into single extension files with {\it splitgroups} and calculate the sky background for each of them with {\it calcsky}. Finally we run {\it dolphot} to calculate the photometry. We use the parameter values recommended for WFC3/UVIS in the user guide. After trying several settings for the sky fitting we find that FitSky=1 and SkipSky=1 give the best results. The counterpart is significantly detected in all frames and has a \g-band magnitude of $27.1 \pm 0.1$ (Vega magnitude system; $27.0 \pm 0.1$ in AB magnitude). This corresponds to an absolute magnitude in the \g-band of $-9.2$ at the distance to SDSS~J1225.

We follow the same procedure with the archival \hst/ACS \g-band observation (data set J8FS34020, taken on 2003 June 16) using the ACS package (version 2.0) of {\sc dolphot}, excluding the J8FS34eqq frame because of a cosmic ray hit at the position of \cxo. We use the parameter values recommended for ACS in the user guide, with FitSky=1 and SkipSky=1. The counterpart is significantly detected at \g = $26.1 \pm 0.1$ (Vega magnitude system; $26.0 \pm 0.1$ in AB magnitude), corresponding to an absolute \g-band magnitude of $-10.2$ at the distance of SDSS~J1225. This is 0.3 mag ($3\sigma$) brighter than the magnitude reported by \citet{jonker10}. This difference is probably due to updated zeropoints and CTE corrections in the {\sc dolphot} ACS package.

We compare the magnitudes of six stars that are in the field of view of both the ACS and WFC3 observations and find that they are consistent with being constant within 2--$\sigma$, with an average difference between the ACS and WFC3 images of $0.07 \pm 0.05$ mag. Hence the difference of $1 \pm 0.1$ mag between the two epochs is not due to our use of different {\sc Dolphot} packages or different settings for the photometric analysis, but to intrinsic variability of the source. 

\section{Discussion}
We have collected new X-ray and optical data of \cxo, reported by \citet{jonker10} as a candidate HLX, recoiling SMBH or extremely X-ray bright SN. From the six new X-ray observations, we learn that the source varies by at least a factor 60 in count rate. After the initial detection in August 2008 at a luminosity of $2.2 \times 10^{41}$ \lum, \cxo{} went undetected in a 10 ks observation in November 2012. In three 20 ks observations in 2014, the source is first barely detected at $7^{+4}_{-5} \times 10^{39}$ \lum{} in April, then not detected in July, then brightens again to be detected at $(5 \pm 2) \times 10^{40}$ \lum{} on November 20. Follow-up observations on December 8 (20 ks) and 15 (40 ks) show a rapid decline of the source luminosity to $< 3 \times 10^{39}$ \lum{} (see Figure \ref{fig:lc}). 

In our new \hst/WFC3 image taken on 2014 June 10 the optical counterpart identified by \citet{jonker10} is still visible, but $1 \pm 0.1$ mag fainter in the \g-band than in the first \hst/ACS image taken in 2003. The \hst{} observations were not simultaneous with X-ray observations. The \hst/ACS observation precedes the first \chan{} observation by five years, so we do not know whether \cxo{} was in outburst or not at that time. In 2014 \chan{} observations were done $\sim 6$ weeks before and after the \hst{} observation. The marginal detection (three counts) in April could either indicate the tail of an outburst that had completely faded at the time of the non-detection in July, or be the result of a random fluctuation of the quiescent flux. In either case it seems reasonable to assume that \cxo{} was in a low state at the time of the second \hst{} observation. If the 2003 observation was taken during an outburst, that would explain why the optical counterpart was brighter at that time than in the 2014 observation.

\subsection{Supernova, recoiling BH or HLX?}
With these new data we can determine which explanation for \cxo{} is the most likely.
The supernova scenario can be discarded --- it is ruled out by the X-ray variability and the fact that the optical counterpart is still visible. The short time-scale of the X-ray variability rules out the possibility that \cxo{} is a background AGN. This was already unlikely because of the blue colour of the counterpart and the high X-ray-to-optical flux ratio of the source (\citealt{jonker10}). 

With the current data we cannot exclude the possibility that \cxo{} is a recoiling massive BH, although --- assuming that the outburst reaches the Eddington luminosity --- the highest luminosity we have detected so far seems rather low for a supermassive BH. The detection of an X-ray source that is positionally consistent with the nucleus of SDSS J1225 casts further doubts on the recoiling BH scenario, as it seems to imply the presence of an SMBH in the centre of the galaxy. However, the luminosity of the X-ray source is low enough to be consistent with a low mass X-ray binary (LMXB) in the nuclear region of SDSS J1225. 

\subsection{Foreground object}
Can \cxo{} be a foreground object? It is located in the direction of the Virgo cluster and the elliptical galaxy NGC 4377 is at a distance of $1.5'$. If \cxo{} is located at the distance of NGC 4377 (18 Mpc; \citealt{2010ApJ...717..603V}), its peak X-ray luminosity would be $\sim 10^{39}$ \lum, and it could be an outbursting LMXB in a globular cluster in the halo of NGC 4377. However, the limit on the absolute magnitude of the optical counterpart argues against this scenario. Based on the 2003 \hst/ACS observations \citet{jonker10} report a limit on the \z-band magnitude of the counterpart of \z{} $> 25.7$. At 18 Mpc, this corresponds to an absolute \z-band magnitude $> -5.5$. This is fainter than any of the globular clusters in the Virgo cluster, and LMXBs are generally detected in the brighter globular clusters with $M_{z'} < -7$ (\citealt{sivakoff07}). 

Another possibility is that \cxo{} is a halo LMXB in NGC 4377 or an intra-cluster LMXB in Virgo, although currently no intra-cluster LMXBs are known. The absolute \g-band magnitude at that distance would be $-5.2$ for the brightest observation. Galactic LMXBs have been observed to reach $M_V \approx -4.0$ during outburst (cf.~\citealt{vanparadijs94}), so this would have to be an unusually bright LMXB.

If \cxo{} were located in the halo of the Milky Way its maximum X-ray luminosity would be $\sim 10^{33}$ \lum{} (assuming a distance of 10 kpc). This could be consistent with a very faint X-ray transient (cf.~\citealt{heinke15}). However, the optical counterpart is more than 3 mag fainter than known optical counterparts to such sources (cf.~\citealt{heinke09}). 

Of the different foreground object scenarios for \cxo{}, a very bright LMXB in the halo of NGC 4377 or in the intra-cluster medium in the Virgo cluster is the only one not ruled out by the data, although it would stretch the parameter space of known LMXBs.

\subsection{The second outbursting HLX}
Taking all these considerations into account, the most likely scenario is that \cxo{} is a bona fide HLX located in SDSS J1225. Its peak luminosity of $2.2 \times 10^{41}$ \lum{} makes it one of the most luminous HLXs known to date. 

For the interpretation of the observations that clearly fall outside the outbursts, i.e.~the non-detections and the detections with three or four counts, one can envisage two scenarios. First, one or both of the detections of a few counts can be seen as evidence for a decaying outburst. Especially for observation 17558 this seems a plausible scenario, as we know that an outburst occurred shortly before that observation. Secondly, all these (non-)detections can be due to random fluctuations of a source that has a luminosity such that the count rate lies just below our detection threshold.

We have seen evidence for two outbursts of \cxo, and the April 2014 detection might indicate the tail of another outburst. If the optical variability is connected to the outbursts, the $\sim 1$ mag brightening in the 2003 \hst{} observation may indicate another one. 
The only other HLX that is known to show outbursts at such high luminosity is HLX-1. In that source, the X-ray flux varies by a factor $\sim 50$, and the $V$-band magnitude of its optical counterpart varies by $\sim 1$ mag over an outburst (\citealt{webb14}). These values are comparable to the ones we find for \cxo. The outbursts of HLX-1 occur with a recurrence time of $\sim 360 - 370$ days, although the last three outbursts were delayed by several weeks to months (\citealt{godet14,kong15atel}). We do not have sufficient data points to detect a recurrence pattern in the outbursts of \cxo.

The duration of the outbursts of \cxo{} is poorly constrained. From our November-December 2014 observations we know that the source decays from $5 \times 10^{40}$ \lum{} to less than $3 \times 10^{39}$ \lum{} in three weeks, but we do not know what the peak luminosity was during that outburst, nor when it started. The quiescent luminosity of \cxo{} is $\lesssim 3 \times 10^{39}$ \lum. This is a factor ten fainter than the luminosity of HLX-1 in its low state (\citealt{servillat11}).

A strong indication for the presence of an IMBH in HLX-1 is the fact that its spectral state changes during its outbursts, in a similar fashion to Galactic LMXBs (\citealt{godet09,servillat11}). Finding similar state changes during outbursts of \cxo{} would confirm it as a strong IMBH candidate. However, in the observations we have obtained so far we do not detect enough counts to constrain the spectral shape during outburst and we have not yet detected \cxo{} at all in quiescence.

\section{Conclusions}
New \chan{} and \hst{} observations of \cxo{} show that the source is most likely a HLX with recurrent outbursts. This is only the second source for which such behaviour has been detected, after HLX-1. \cxo{} is less luminous than HLX-1 and decays on a shorter time-scale. More observations are needed to determine whether its outbursts show a recurrence pattern similar to HLX-1.
The characteristics of \cxo{} make it a very interesting candidate IMBH. 

\section*{Acknowledgements}
We would like to thank T. Maccarone for interesting discussions and the anonymous referee for their useful and detailed comments that helped improve the Letter. The results reported in this Letter are based on observations made by the \chan{} X-ray Observatory and the {\it Hubble Space Telescope}. We thank the \chan{} X-ray Center's director Belinda Wilkes for granting us Director's Discretionary Time observations of \cxo. 

\bibliographystyle{mn_new}
\bibliography{bibliography}

\end{document}